\documentstyle[aps,multicol,eqsecnum,epsf,rotate]{revtex}

\begin{document}
\draft
\tighten

\def\bfr{{\bf r}}
\def\vecr{{\vec r}}
\def\lim{\mathop{\bf\rm Limit}}
\def\tphi{\partial_t\tilde\phi}
\def\grad{\pmb{$\nabla$}}
\def\Lx{L_x}

\title{Dynamical Transition in Sliding Charge-density Waves with
Quenched Disorder}

\author{Lee-Wen Chen$^{(1)}$,
        Leon Balents$^{(2)}$, Matthew P. A. Fisher$^{(2)}$\\
        and M. Cristina Marchetti$^{(1)}$\\}
\address{
$^{(1)}$  Department of Physics, Syracuse University\\
$^{(2)}$  Institute for Theoretical Physics, University of
California, Santa Barbara}
\date{\today}
\maketitle
\begin{abstract}
\noindent
We have studied numerically the dynamics of sliding charge-density
waves (CDWs) in the presence of impurities in d=1,2. 
The model considered exhibits a first order dynamical
transition at a critical driving force $F_c$ between  ``rough'' 
(disorder dominated) $(F<F_c)$ and ``flat'' $(F>F_c)$
sliding phases where disorder is washed out by the external drive.
The effective model for the sliding CDWs in the presence of impurities
can be mapped onto that of a magnetic flux line pinned by columnar
defects and tilted by an applied field. The dynamical transition of
sliding CDWs corresponds to the transverse Meissner effect of the
tilted flux line.
\end{abstract}
\pacs{PACS numbers: 71.45.Lr, 72.70.+m, 74.60.Ge }

\begin{multicols}{2}

\bibliographystyle{plane}
\noindent
\section{Introduction}
The study of the dynamics of an ordered medium driven by an external
force through quenched impurities is relevant to many physical
systems.  Examples include charge-density waves (CDWs) in anisotropic
metals\cite{gruner} and flux-line lattices (FLL) in the mixed state of
type-II superconductors\cite{fisherhuse}.  It has been argued that in
both these systems impurities destroy the equilibrium long-range
translational order (LRTO) and pin the medium. A driving force
$F$, originating from an electric field or a current, can
overcome the constraining forces from the impurities and cause the
medium to slide. At $T=0$, classical CDWs exhibit a depinning
transition at a critical
driving force $F_T$ from a pinned ($F<F_T$) to a
sliding ($F>F_T$) state. This transition has been described
as a dynamic critical phenomenon. The nonlinear dynamics of the system
near the critical point has been studied extensively both by numerical
simulation\cite{cdwsimul} and by $4-\epsilon$
expansions\cite{cdwnarayan} and is fairly well understood.

In contrast, the dynamics of driven disordered media at large driving
force, well above the depinning transition, has only recently begun to
receive some attention. In the sliding state the pinning by impurities
is less effective and it has been suggested that the medium may
recover the LRTO at sufficiently large velocity. Recent experiments in
YBCO as well as simulations of two-dimensional flux lattices have
indeed shown that the flux array orders at large drives. Koshelev and
Vinokur \cite{vinokurmoving} have described this phenomenon as a true
phase transition from a flowing liquid to a moving solid. Whether the
ordering of driven flux lattices is a true dynamical phase transition
or a crossover is still an open question.

Some of us recently addressed this class of questions by focusing on a
model of driven CDWs in the presence of disorder and thermal
noise\cite{balents}. The model considered in Ref.\cite{balents}\
allows for disorder-induced phase slips of the CDW. In three
dimensions it yields a dynamical phase transition of the sliding CDW
from a disordered phase with plastic flow to a temporally periodic
``moving solid'' phase with quasi-long-range translational order. In
two dimensions the moving solid phase is unstable due to the
proliferation of phase slips.

In the present paper we focus on a related model of the dynamics of
sliding CDWs that incorporates a nonequilibrium nonlinear term of the
Kardar-Parisi-Zhang (KPZ) form\cite{kpz}, that was neglected in
Ref.\cite{balents}.  To render the analysis tractable with this
addition, we have, however, neglected phase slips, and, for
most of what follows, thermal fluctuations.  While the neglect of
thermal noise has very little effect upon our results, the omission of
phase slips appears more severe.  Indeed, the model considered here is by
definition an elastic continuum at all driving forces and a ``liquid
phase'' with dislocations cannot occur.  We expect that at
sufficiently low temperature and large driving force the omitted
defects will make only minor changes to our results, at least on
experimentally observable time scales.  The true (infinite time)
asymptotics is, however, likely to be affected by this omission.

Interestingly, despite its strong topological constraints, our model
nevertheless exhibits a phase transition at a critical driving force
$F_c$ from an isotropic rough (disorder dominated) flowing phase
at small driving forces, to an anisotropic smooth flowing phase where
nonlinearities are washed out by the external drive.  The average CDW
velocity changes sharply at the transition, which is argued to be
first order. The two phases are characterized by different values of
the roughening exponent governing the growth of spatial fluctuations
of the CDW phase with the size $L_x$ of the system in the direction of
the external drive ($x-$direction).  In the rough phase the phase
fluctuations grow linearly with $L_x$, $w(L_x,L_\perp)\sim L_x$,
indicating that the elastic model breaks down. This suggests that in a
corresponding model that allows for phase slips, this phase would be a
flowing liquid. Above $F_c$ the disorder is washed out in the
direction of motion.  Fluctuations are strongly suppressed in this
direction and $w(L_x,L_\perp)\sim L_x^{1/2}$ for $L_x^z>>L_\perp$, with
$z\approx 0.85\pm 0.05$.

Our results have two implications for CDW experiments at high
velocities, provided phase slip effects are sufficiently suppressed at
low temperatures.  First, above the critical force
($F>F_c$), translational correlations are expected to be
highly anisotropic, decaying much more rapidly transverse to the
motion than along it.  This implies a substantial increase in the
ratio of widths of Bragg scattering peaks,
\begin{equation}
{d \over {dF}} \left( {{\Delta k_\perp} \over {\Delta k_\parallel}}
\right) > 0,
\label{peakwidths}
\end{equation}
where $\Delta k_\perp$ and $\Delta k_\parallel$ are peak widths
perpendicular and parallel to the CDW wave vector, respectively.
Secondly, within our model, the non-equilibrium ordering transition is
characterized by a jump discontinuity in the differential conductance
$G_{\rm diff} \equiv dI/dV$.  As discussed in further detail in the
next section, we are able to predict only the singular behavior of
$G_{\rm diff}$, and not the full non-linear form of the $I(V)$ curve.

We expect these two results to survive in varying degrees in models
including phase slips (and hence experiments).  The strong anisotropy
of the translational correlations in the high-velocity phase should
remain at low temperatures.  The resistance singularity is expected to
be more sensitive to phase slips (as, indeed, is the entire small
$F$ phase).  Their effects are expected to round the step in
$G_{\rm diff}$ very near to $F_c$.

\section{Effective model for sliding CDWs}
Charge-density waves are coupled electron-phonon excitations which
exist in a class of anisotropic metals consisting of weakly coupled
chains. In these materials the electronic density is sinusoidally
modulated along the chain $(x-)$ direction,
\begin{equation}
\rho(\bfr)=\rho_0+\rho_1\cos(2k_Fx+\phi(\bfr,t)),
\label{cdw}
\end{equation}
where $k_F$ is the in-chain Fermi wave vector and $\rho_1$ the
amplitude of the charge modulation. At low temperature, due to the gap
in the dispersion relation for the amplitude, amplitude fluctuations
are strongly suppressed and the dynamics can be described in terms of
the phase $\phi$ only.  The Hamiltonian $\cal H$ for a CDW in a
$d-$dimensional metal can be written as
\cite{gruner}
\begin{eqnarray}
{\cal H} & = &{K \over 2}\int d{\bf r}\big[(\nabla\phi)^2\Big] 
\nonumber \\ && + \int
d\bfr V({\bf r})\rho_1\cos(2k_Fx+\phi({\bfr},t)), 
\label{cdwh2d}
\end{eqnarray}
where we have rescaled coordinates to obtain an isotropic elastic
term.  The coefficient $K$ is a stiffness constant.  The effect of
impurities is described via a Gaussian random potential $V(\bfr)$ with
zero mean, $<V(\bfr)>=0$, and short range correlations,
$<V(\bfr)V({\bf 0})>=V_0^2\Delta(|\bfr|/\xi_0)$, with $\xi_0$ a short
wavelength cutoff.  The overdamped equation of motion for the CDW
phase variable $\phi$ is given by\cite{flr},
\begin{eqnarray}
\partial_t\phi&=&-{D \over K}{\delta{\cal H}\over\delta\phi} + \omega_0
\nonumber\\
&=&D\nabla^2\phi+\omega_0 +\tilde V(\bfr)\sin(2 k_Fx+\phi(\bfr,t)),
\label{flreom}
\end{eqnarray}
where $D=({m\over m^*})\tau v_F^2$ has the dimension of a diffusion
constant $(l^2t^{-1})$ .  Here $\tau$ is the relaxation time of a
thermally excited phonon, while $m$ and $m^*$ are the electronic mass
and effective mass, respectively.  We have also let $\tilde
V(\bfr)=(D/K)V(\bfr)$.  The second term on the right hand side of
Eq. (\ref{flreom}) arises from an electric field $E$ applied along the
$x-$direction, with $\omega_0=Ee({\tau\over m^*})2k_F$, and can be
shifted away by $\phi\rightarrow\phi+ \omega_0 t$.  The ``force''
$\omega_0$ has dimensions of frequency $(t^{-1})$ and actually
represents the ``washboard frequency'' $\omega_0 = 2k_F v$ of a freely
sliding CDW (with velocity $v$) driven by an external electric field
$E$, in the absence of quenched disorder.  Equation (\ref{flreom}) is
the conventional Fukuyama-Lee-Rice (FLR) model of CDW
dynamics\cite{flr}, which has been studied extensively both
analytically and numerically, particularly near the depinning
transition.  The FLR model exhibits a depinning transition at a
threshold field $E_T$, corresponding to a threshold force
$\sigma_T=E_Te({\tau\over m^*})$.

As discussed in Ref.\cite{balents}, the FLR equation is
incomplete in the strongly driven regime.  It is essentially
a near-equilibrium description, in which only the most relevant
perturbation (the driving field) has been added to the
equilibrium relaxational dynamics.  Several additional effects become
important in the sliding state.

The most important such effect is that of {\sl convection}.  In
particular, in a CDW moving with velocity $v$, the partial time
derivative $\partial_t$ in Eq.~\ref{flreom}\ must be replaced by the
total convective derivative $D_t = \partial_t + v \partial_x$.  More
generally, the linear derivative ($\partial_x\phi$) term arises
because the electric field breaks the reflection symmetry $x
\rightarrow -x$.  Note that the coefficient of this term is small for
small velocities, which is why it is neglected in the usual
equilibrium and near-static (i.e. CDW depinning) contexts.

A second term ordinarily omitted from the FLR equation in equilibrium
arises from coupling to the underlying periodic lattice.  This
intrinsic pinning in the direction of motion can be incorporated in
Eq. (\ref{flreom}) by the replacement $ V(\bfr )\rightarrow V(\bfr
)+W(x)$, where $W(x)$ is a periodic potential, $W(x)=W_0\cos(Qx)$, and
$Q$ is in general incommensurate with $2 k_F$.  While in other
contexts such an incommensurate periodic potential can be safely
neglected, we will see that it gives rise to important effects for the
asymptotic behavior in the strongly driven limit.

Including both these effects, we arrive at a suitable generalization
of the FLR equation,
\begin{eqnarray}
\partial_t\phi & = & D\nabla^2\phi+\omega_0 - \sigma\partial_x\phi
\nonumber \\ & & +\big[\tilde V(\bfr)+\tilde W(x)\big]\sin(2k_F 
x+\phi(\bfr,t)),
\label{generalflr}
\end{eqnarray}
where again a numerical factor has been absorbed into the periodic
potential $\tilde W(x)$.  We have allowed for renormalizations of the
convective term by keeping the coefficient $\sigma$ general, but we
expect $\sigma \sim v$.  Eq.~\ref{generalflr}\ is capable of
describing the behavior of the CDW (up to the aforementioned caveats
respecting phase slips and thermal fluctuations) in the full range of
applied fields from well below to far above the nominal threshold
field.

In fact, Eq.~\ref{generalflr}\ is so general that it is a rather
inappropriate point from which to study the moving state.  This is
made evident by making the transformation $\phi = \omega_0 t +
\tilde{\phi}$, in order to focus on the fluctuations $\tilde{\phi}$
around the uniformly sliding CDW.  The resulting equation of motion
for $\tilde{\phi}$ contains force terms which oscillate rapidly in
time.  To determine their effect at time scales longer than
$2\pi/\omega_0$, one must develop instead an {\sl effective} equation
of motion for a coarse-grained (temporally and spatially averaged)
phase $\overline{\tilde{\phi}}$.  In what follows we will drop the
overbar and denote the coarse-grained phase variations simply by
$\tilde{\phi}$.  

The coarse-graining procedure may be explicitly performed in two
different ways.  The simplest method is a variant of the high-velocity
expansion about the sliding state\cite{DSFcdw}, obtained by iterating
a formulation of Eq.~\ref{generalflr}\ as an integral equation.  A
more complicated, but conceptually more clear approach is to
coarse-grain using renormalization group (RG) methods, in which short
wavelength and high frequency components of $\tilde{\phi}$ are
explicitly integrated out in a field-theoretic formulation.  A similar
calculation was carried out recently in a different context by Rost
and Spohn\cite{sinegordonrg}.  Both approaches are straightforward but
tedious, and we simply quote the results in what
follows\cite{unpublished}.

Several simplifications are obtained in this effective coarse-grained
description.  The most important is the modification of the random
potential term $\tilde V(\bfr) \sin(2k_F x+\phi)$, which, as mentioned
before becomes oscillatory.  A careful treatment reveals, however,
that this term does not strictly average to zero in the coarse-grained
model.  Instead, as discussed in Refs.\cite{krug,balents}, it
generates an effective spatially varying drag force $F_p(\bfr )$.
To leading order in $1\over\omega_0$, its correlations are 
$<F_p(\bfr)F_p({\bf0})>=F_0^2\delta(\bfr)$ 
with $F_0={\tilde V_0^2\over 4\omega_0}$.  This may be understood
physically as simply reflecting variations of the impurity density in
different regions of the sample, which then exert a spatially random
drag force on the CDW.  

An important difference between this term and the original sine-Gordon
type term is that it does not prefer any particular value of the phase
variation $\tilde{\phi}$.  This is in fact an exact result in the
moving phase, reflecting the non-trivial transformation property
$\tilde{\phi} \rightarrow \tilde{\phi} +
\omega_0 \tau$ under a time-translation $t \rightarrow t + \tau$.  In
general terms, the equilibrium ordered phase of the CDW is described
as a state of spontaneously broken spatial translation symmetry.  This
state is highly susceptible to disorder, because randomness explicitly
breaks precisely this symmetry -- i.e. it acts as a random field.  By
contrast, the sliding CDW breaks {\sl time}-translation symmetry,
which is an {\sl exact} invariance of the system, even with $\tilde{V}
\neq 0$.

A second simplification occurs in the intrinsic pinning term.  Like
the random potential, this term also becomes oscillatory in time, but
generates a non-trivial correction upon coarse-graining.  To second
order in a gradient expansion, the correction has the form of an
additional drag force $\delta F_{W} \sim -{W_0^2\over 2\omega_0}[1 -
c_1 |\nabla_\perp\tilde{\phi}|^2 - c_2 |\partial_x\tilde{\phi}|^2]$,
where $c_1 \sim c_2 \sim 1/(2k_F)^2$ are constants.  Physically, the
gradient corrections arise because the drag force from intrinsic
pinning becomes less effective as the CDW wavevector (whose local
shift is proportional to $\nabla\tilde{\phi}$) becomes less
commensurate with the underlying lattice.  For simplicity, we will
focus on the isotropic case $c_1 = c_2$, which is expected to be
approximately correct for CDWs whose density profile is well
approximated by the single Fourier harmonic form of Eq.~\ref{cdw}\ and
which is not too far from commensurability.  The resulting
gradient-squared correction is a realization of the
Kardar-Parisi-Zhang (KPZ) nonlinearity in the CDW system.

The final coarse-grained equation of motion is 
\begin{equation}
\partial_t\tilde\phi=D\nabla^2\tilde\phi-\sigma\partial_x\tilde\phi+F_p(\bfr)
+{\lambda_0\over 2}(\nabla\tilde\phi)^2,
\label{eomeffective}
\end{equation}
where $\lambda_0 \sim {W_0^2\over {(8\omega_0 k_F^2)}}$.  This
coefficient is positive, because a mis-oriented CDW (with
$\nabla\tilde\phi \neq 0$) is less slowed
down than an aligned one (with $\nabla\tilde\phi = 0$).

Eq.~\ref{eomeffective}\ is the basis for our study of the moving
state.  We caution, however, that some information is lost in this
approach, and various non-universal high energy features of the CDW
behavior are no longer easily calculable.  An important example is the
full form of the I-V curve.  As can be explicitly seen in the
coarse-graining procedure, the CDW frequency $\partial_t \phi$ as a
function of $E$ or $\omega_0$ has non-trivial contributions from the
short-wavelength degrees of freedom not contained in
Eq.~\ref{eomeffective}.  An additional difficulty is that the drag
forces $F_0$ and $\lambda_0$ are strongly force dependent.  Our
long-wavelength description {\sl does}, however, capture the {\sl
singular} part of the CDW velocity.  We define 
\begin{equation}
\delta v_{sing} =
\overline{<\partial_t\tilde{\phi}>},
\label{velocity}
\end{equation}
where the overbar denotes a spatial average and the brackets denote
the disorder average.  The quantity $\delta v_{sing}$ 
is actually a frequency
shift.  Note that, because this includes only the singular part of the
$v(\omega_0)$ relation, there is no particular preferred sign for
$d\delta v_{sing}/d\omega_0$.  

The spatial fluctuations of the phase can be
characterized by their growth with the system size.
A useful measure of such fluctuations employed in the study
of interface dynamics is the "interface width" in the long-time
saturated regime, given by
\begin{equation}
w(L_x,L_\perp)\equiv
<\overline{[\phi(\bfr,t)-\overline{\phi(\bfr,t)}]^2}>^{1/2}
\label{width}
\end{equation}
in a d-dimensional system of size $L_xL_\perp^{d-1}$. 

If the KPZ term
is neglected in Eq. (\ref{eomeffective}), the equation is linear and
can be solved exactly by Fourier transformation, as discussed in
Ref.\cite{balents}. The CDW response is linear and
$<\overline{\partial_t\phi}>=\omega_0$, i.e., $\delta v_{sing}$=0. 
The random mobility yields a
static distortion of the CDW,
\begin{equation}
\tilde\phi({\bf q},\omega)={F_p({\bf q})\over
Dq^2+i\sigma q_x}2\pi\delta(\omega).
\label{fourierresponse}
\end{equation}
The corresponding correlation function is $<|\tilde\phi({\bf
q,\omega})|^2>=S(q)2\pi\delta(\omega)$, with $S(q)={F_0^2\over
(Dq^2)^2+\sigma^2q_x^2}$ the static structure function. The
$\sigma\partial_x\phi$ term in Eq. (\ref{eomeffective}) is crucial in
determining the decay of spatial correlation in the moving state. If
this term is absent, fluctuations are isotropic, with $S(q)\sim
q^{-4}$, so that $w(L)\sim L^{(4-d)/2}$, for $L_x=L_\perp=L$. In
particular in $d=1$, $w(L)\sim L^{3/2}$ and the system will develop a
``groove'' instability of the type discussed in \cite{family}. The
case $d=2$ is marginal with $w(L)\sim L$. 
The $\sigma\partial_x\phi$ term
suppresses the growth of fluctuations in the $x-$direction. When
this term is present, in the limit where
$L_x>>L_\perp$ and $\sigma>>1$, the CDW is ``riding over'' the static
disorder and $w(L_x,L_\perp)\sim L_x^{1/2}$.

\section{Tilted magnetic flux line analogy}
When the KPZ coupling $\lambda_0$ is nonzero, Eq. (\ref{eomeffective})
can be mapped into the problem of a directed path in a random
potential via the well known Cole-Hopf
transformation\cite{etras}. By letting 
$\Psi(\bfr,t)=e^{{\lambda_0\over 2D}\tilde\phi(\bfr,t)},$ a linear
equation of motion for $\Psi(\bfr,t)$ is obtained,
\begin{equation}
\partial_t\Psi=[\sigma\partial_x+D\nabla^2+{\lambda_0\over 2D}F_p(\bfr)]\Psi.
\label{burger}
\end{equation}
The solution of Eq. (\ref{burger}) can be written as a
path-integral,
\begin{equation}
\Psi(\bfr,t)=\int\limits_{({\bf 0},0)}^{(\bfr,t)}{\cal D}[\bfr]
e^{{-1\over 2D}\int\limits_0^tdt'[{1\over
2}(d\bfr/dt')^2-\sigma\hat{x}-\lambda_0 F_p(\bfr)]}.
\label{pathint}
\end{equation}
Eq. (\ref{pathint}) can also be interpreted as
the partition function of a tilted magnetic flux line in the presence
of columnar pinning centers. A single magnetic flux line 
in a $(d+1)-$dimensional sample of
thickness $L$ in the direction of the applied field ${\bf H}$, chosen
as the $z$ direction $({\bf H}=H_0\hat{\bf z})$, is 
parametrized by its
trajectory ${\bf r}(z)$ as it traverses the sample along the field
direction. The sample contains columnar pinning centers aligned with
the $z-$direction that can pin the flux line over its entire length.
An additional magnetic field ${\bf H}_\perp$ applied perpendicular to
the $z-$direction tilts the flux line away from the direction of the
columnar pins.
The flux line free energy is then given by,
\begin{equation}
G=\int
\limits_0^Ldz\Big\{{\tilde\epsilon_1\over 2}\Big[{d{\bf r}\over
dz}-{{\bf h}\over\tilde\epsilon_1}\Big]^2+U({\bf r}(z))\Big\},
\label{lineenergy}
\end{equation}
where $\tilde\epsilon_1=({\phi_0\over 4\pi\lambda_{ab}})^2ln(\kappa)$
is the tilt 
modulus (we assume for simplicity an isotropic superconductor),
${\bf h}={\bf H}_\perp\phi_0/4\pi$, and
$U({\bf r})$ is the random pinning potential generated by the columnar
defects. The pinning potential is correlated along the direction and
has short range correlations in the plane, with $<U({\bf r})U(\bfr')>
=\Delta\delta({\bf r}-{\bf r}')$.
The partition function of a vortex line with fixed end points ${\bf
r}(0)={\bf 0}$ and ${\bf r}(L)={\bf r}_\perp$ is obtained by
summing the Boltzmann factor $e^{-G/k_BT}$ over all paths connecting
the end points and it is given by
\begin{equation}
Z=\int\limits_{({\bf 0},0)}^{(\bfr,t)}{\cal D}[\bfr]
e^{-{1\over T}\int\limits_0^Ldz\Big\{{\tilde\epsilon_1\over 2}
\Big[{d{\bf r}\over
dz}-{{\bf h}\over\tilde\epsilon_1}\Big]^2+U({\bf r}(z))\Big\}}.
\label{partition}
\end{equation}
The dynamics of a $d-$ dimensional driven CDW at $T=0$ can 
therefore be mapped onto 
the tilt response of a magnetic flux line in a $(d+1)-$dimensional
superconductor with columnar pins,
at finite temperature. In this mapping the time argument of the CDW
corresponds to the flux line coordinate $z$ along the field direction,
the diffusion constant $D$ plays the role of temperature, according to
$D\rightarrow {T\over 2\tilde\epsilon_1}$, and the driving force
$\sigma$ corresponds to a tilt field ${\bf
h}={h}\hat{x}$ with $\sigma\rightarrow{h\over\tilde\epsilon_1}$. The
correspondence between the various CDW and flux line quantities is
summarized in table 1. 

\begin{center}
\def\arraystretch{1.5}
\begin{tabular}{|c|c|}\hline
${\rm\qquad CDW\qquad\qquad}$	& ${\rm\qquad Flux\quad line\qquad}$ \\ \hline\hline
$ D$	& $ T/2\tilde\epsilon_1$ \\ \hline
$\sigma$ & $h/\tilde\epsilon_1$  \\ \hline
$\lambda_0F_0$ & $\sqrt\Delta/\tilde\epsilon_1$ \\ \hline
$-\lambda_0\delta v_{sing}$ & $g/\tilde\epsilon_1$\\ \hline
$\lambda_0{d\delta v_{sing}\over d\sigma}$ & $m_\perp$ \\ \hline
\end{tabular}
\vspace{20pt}
\centerline{\bf Table 1}
\end{center}

Eqs. (\ref{partition}) and (\ref{pathint})
differ by a constant term $h^2L\over 2\tilde\epsilon_1$ in the flux
line free energy which represents the field energy associated with the
tilt field ${\bf H}_\perp$. In the absence of a tilt field, the flux
line is localized on the strongest columnar defect. At low temperature
the localization length, defined as the radius of the tube to which
the flux line is confined, is of the order of the range of the
pinning potential. Thermal fluctuations increase the localization
length, but are not sufficient to depin the flux line in $d=1,2$. A
sufficiently strong perpendicular field ${\bf H}_\perp$ will, however,
depin the flux line. The response of the flux line to the field
is measured by the average induction in the direction $B_\perp$
of the transverse field. 
We define a dimensionless induction 
$b_\perp={B_\perp\over n_0\phi_0}$. This is also the mean slope of the
tilted flux line. The induction can be written as
$b_\perp={h\over\tilde\epsilon_1}+4\pi m_\perp$, where
$m_\perp=-{\partial g\over \partial h}$ is the (dimensionless) total
magnetization and $g(h)$ is the Gibbs free energy per unit length of
the tilted flux line. It has been argued that a flux-line array
pinned by columnar defects exhibits a transverse Meissner effect, with
$b_\perp=0$ for tilt field below a critical value $h_c$ and
$b_\perp\neq 0$ for $h>h_c$. The tilt response of a {\it single} flux
line pinned by {\it one} columnar defect in $(1+1)-$dimensions
can be evaluated analytically (see Appendix A and \cite{hatano}). 
One finds that
in the limit $L_x\rightarrow\infty$ there is a transverse Meissner
effect for $h<h_c$. For a $\delta-$function pin with
$U(\bfr)=-U_0\delta(\bfr)$, we find 
$h_c={U_0\over T}{4\pi\tilde\epsilon_1\over\phi_0}$. 
In the pinned configuration for $h<h_c$ the flux line free
energy per unit length is $g={(h^2-h_c^2)\over 2\tilde\epsilon_1}$, so that
$m_\perp=-{h\over\tilde\epsilon_1}$ and $b_\perp=0$.
For $h\geq h_c$ the line is depinned and $g=0$. This gives $m_\perp=0$
and $b_\perp={h_\perp\over\tilde\epsilon_1}$.
The transition from a pinned to a depinned
configuration is associated with a jump discontinuity in the induction
or tilt slope $b_\perp$ at $h_c$ and can therefore be classified as a
{\it first order} phase transition. For a general pinning potential we
estimate $h_c$ as the field required to depin the flux line,
$h_c^2/2\tilde\epsilon_1\sim\sqrt\Delta$, or
$h_c\sim(2\tilde\epsilon_1{\sqrt\Delta})^{1/2}$. 
Similar conclusions were reached
by Balents and Simon \cite{balentssimon} for the tilt response of a
single flux line in a random distribution of columnar pins in
$(1+1)-$dimension. Also Hatano and Nelson \cite{hatano} very recently
related the depinning of a flux line from columnar defects by a
transverse field to the localization transition of a quantum particle
in a constant imaginary vector potential. By exploiting this mapping
they showed that the transverse Meissner effect persists in both
$d=1,2$. The question of whether an array of many {\it interacting}
flux lines pinned by columnar defects will also exhibits a sharp
transition is, however, still open.

The transverse Meissner effect for the vortex line translates into a
{\it first order} dynamical phase transition of the sliding CDW. We
recall that the CDW driving force $\sigma$ corresponds to the tilt
field $h$ and $\delta v_{sing}\sim g$.
There is then a transition at a characteristic $\sigma_c$
from a state with $\delta v_{sing}\neq 0$ for $\sigma<\sigma_c$ to a
state where the external drive dominates and washes out the effect of
disorder in the direction of the drive, yielding $\delta
v_{sing}=0$ for $\sigma<\sigma_c$. The driving force $\sigma_c$ 
where the transition occurs can be
estimated from the flux-line analogy using table 1 and the estimated
$h_c\approx(2\tilde\epsilon_1{\sqrt\Delta})^{1/2}$
as $\sigma_c\approx\sqrt{2\lambda_0
F_0}$. The sliding phase with $\delta v_{sing}\neq 0$ at small
driving forces corresponds to the situation where the flux line is
pinned on the strongest columnar defects and exhibits a transverse
Meissner effect. As we will show in the next section, this is a
disorder-dominated regime for the sliding CDW with a ``rough'' spatial
profile of the phase $\phi({\bf r}, t)$. We will refer to this phase as
a rough sliding phase. For $\sigma\geq\sigma_c$ the CDW is in a
sliding phase with $\delta v_{sing}=0$, corresponding to a flux line
depinned by the tilt field and ``riding over'' the columnar pins. As
shown below, this phase is characterized by anisotropic spatial
fluctuations of the phase. Fluctuations are suppressed in the
direction of the driving force and we will refer to this phase as a
``flat'' phase.

\section{Numerical results} 
We have integrated numerically Eq. (\ref{eomeffective}) in
both $d=1,2$ by discretizing
the spatial coordinates, with
lattice spacing equal to the range $R_p$ of the pinning
potential, chosen as our unit of length. 
We assume an initially flat configuration
, $\phi(\bfr,t=0)=0$, and follow the dynamics until
the system relaxes to a steady state. The
relaxation time scales as $L_x^{2}L_\perp^{2d-2}$. 
The average CDW properties in the saturated sliding
state are evaluated by performing both a time average and an average
over many realizations of the disorder.
\begin{figure}[hbt]
\epsfxsize=3.5in\epsfbox{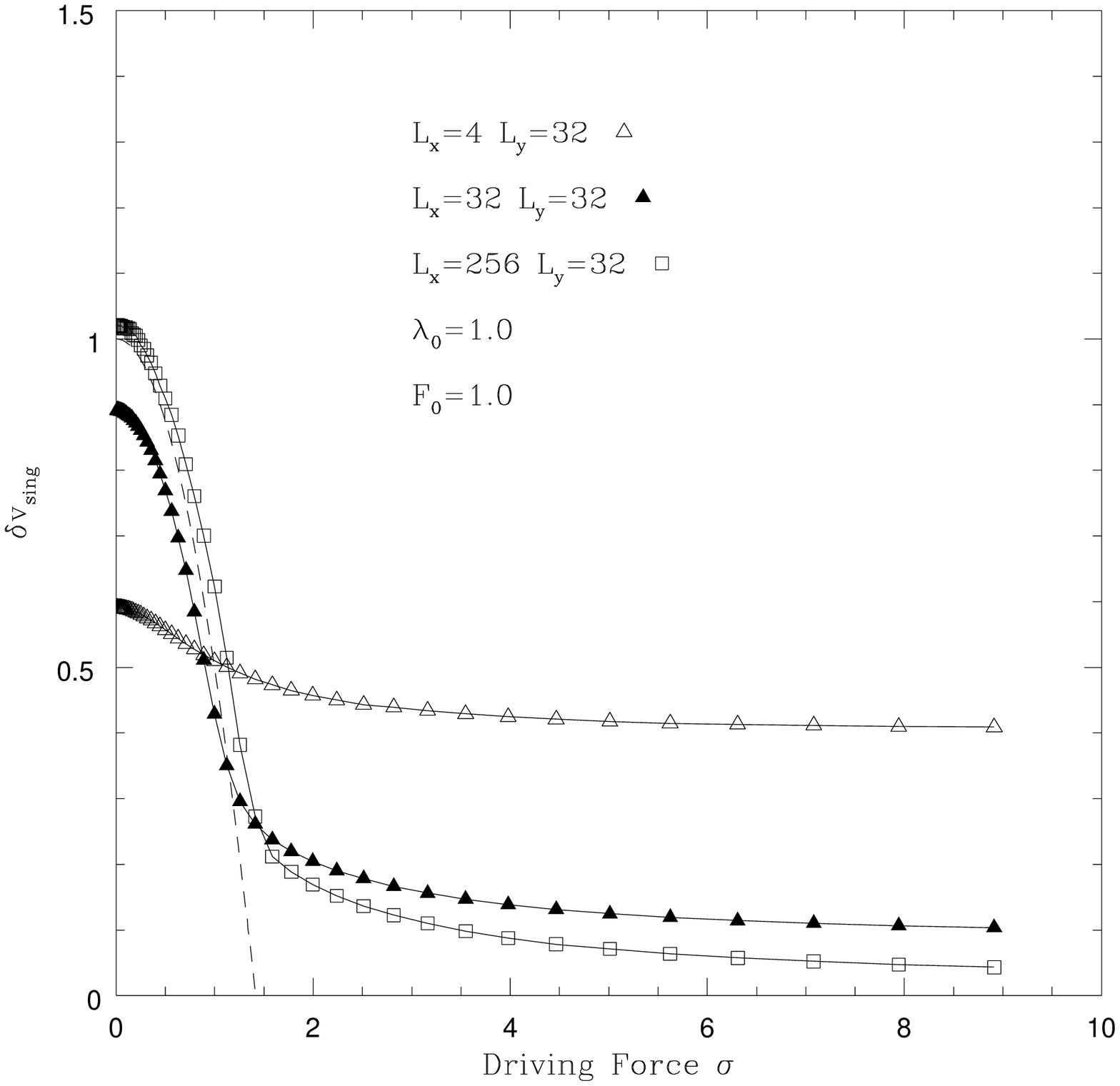}
\vspace{5pt}
\centerline{\bf Fig.1}
\end{figure}

Figure 1 shows the frequency shift $\delta v_{sing}$ of the CDW
defined in Eq. (\ref{velocity}) as a function of the
driving force $\sigma$ for $d=2$. 
A similar behavior is obtained in $d=1$. As
the system size $L_x$ is increased one observes a transition between
two sliding phases. For the parameter values used here the estimated
critical force is $\sigma_c=\sqrt 2$, a value that agrees quite well
with our numerical results. At large driving forces disorder is washed
out by the external drive and $\delta v_{sing}$ approaches zero as
$L_x\rightarrow\infty$. For small driving forces both impurities and
intrinsic pinning are important and yield a large $\delta v_{sing}$.
While the drop of $\delta v_{sing}$ above a critical force does become
sharper as $L_x$ increases, the approach to the sharp transition
expected in the limit $L_x,L_\perp\rightarrow\infty$ is rather slow.
This can be understood by examining the dependence on the system size
$L_x$ of the free energy of a single flux line pinned by a single
columnar defect evaluated in Appendix A. As discussed in the previous
section, this simple model exhibits a first-order depinning transition
in the limit $L_x\rightarrow\infty$. On the other hand, the finite size
corrections to the flux line free energy are large in the region
$h> h_c$, as shown in Fig. 5. For $h<h_c$ the flux line is localized
on the columnar pin and does not ``see'' the rest of the system. As a
result, in this region the finite size corrections to the free energy
vanish exponentially with system size. For $h>h_c$ the flux line is
delocalized and samples the entire system. In this region the free
energy is quite sensitive to the finite system size, with
$g(L_x)-g(\infty)\sim {1\over L_x}$. 
The scaling of $\delta v_{sing}(L_x,L_\perp)$ in our driven CDW
problem - which
maps onto a tilted flux line pinned by a random distribution of {\it many}
columnar defects -- is even slower than obtained in the single pin
model , with $\delta
v_{sing}\sim L_x^{1/2}$ approximately. This slow approach to the
asymptotic limit can, however, be understood by the same physical
argument.  

\begin{figure}[hbt]
\epsfxsize=3.5in\epsfbox{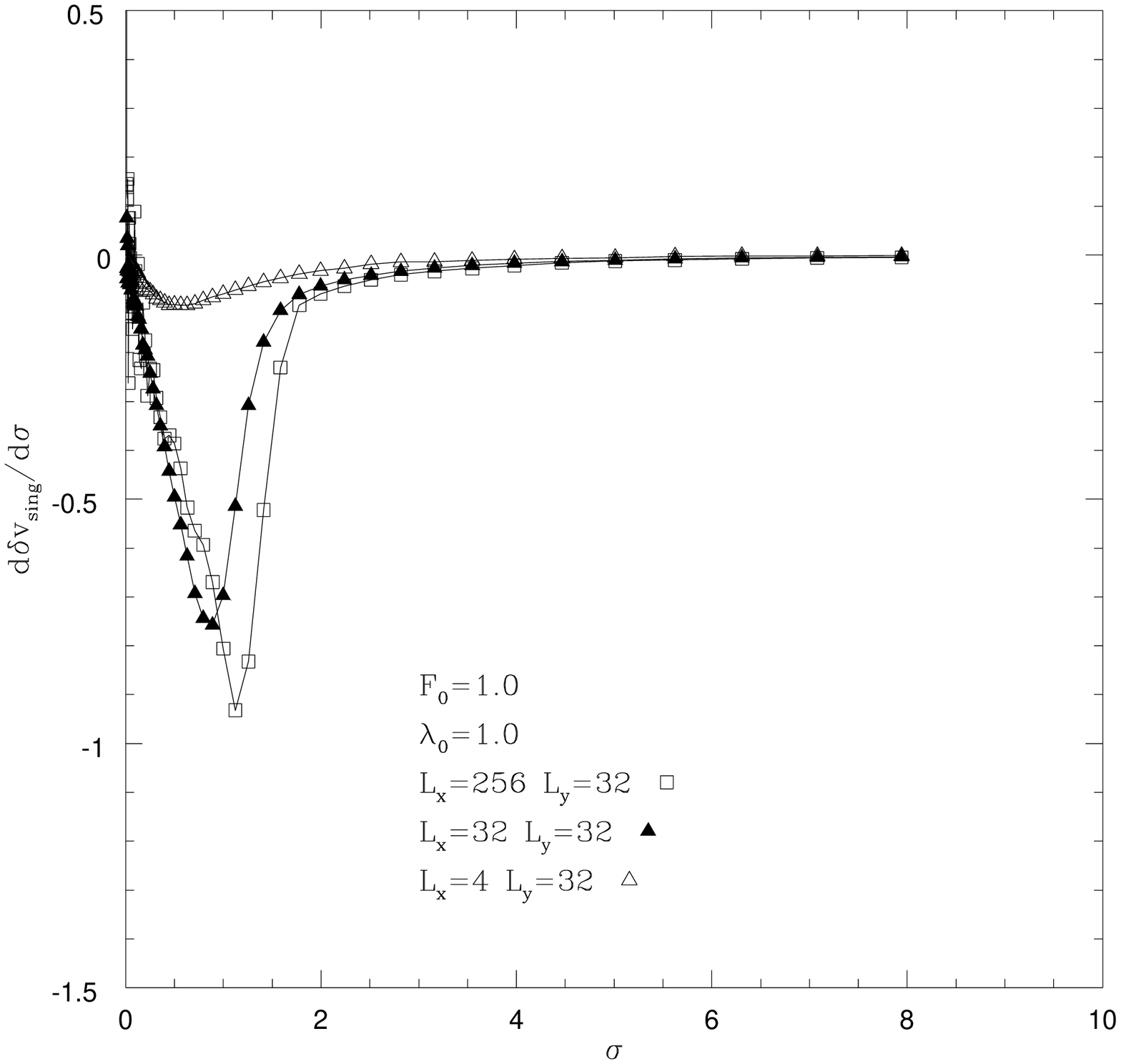}
\vspace{5pt}
\centerline{\bf Fig.2}
\end{figure}

To better display the change of $\delta v_{sing}$ at the transition we
show in Fig. 2 the evolution of the curve $d v_{sing}\over
d\sigma$ versus $\sigma$ with system size $L_x$ for $d=2$. 
In the limit $L_x\rightarrow\infty$ the
derivative will exhibit a jump discontinuity at the transition.
As the driving force $\sigma$ is proportional to the applied voltage
(electric field) and $\delta v_{sing}$ is proportional to the current $I$,
this corresponds to a jump
discontinuity in the differential conductance $G_{\rm diff}={dI\over dV}$. 
Our model only allows us to predict the singular
shift $\delta v_{sing}$ and therefore the singular behavior of $G_{\rm
diff}$, but not the full nonlinear form of the $I(V)$ curve. The
magnitude of the jump discontinuity in $dv_{sing}\over
d\sigma$ is, however, proportional to the jump discontinuity in
$G_{\rm diff}$. The precise relation and the possibility of observing
this effect will be discussed in the next section.
From table 1 we see that $d v_{sing}\over d\sigma$ corresponds to
the magnetization of the tilted flux line. Fig. 2 displays then the
transverse Meissner effect discussed earlier.

By translating the results obtained in Appendix A for the flux line,
we find that if the random force $F_p({\bf r})$ of Eq. (\ref{burger})
is replaced by a single $\delta-$function pin, the frequency $\delta
v_{sing}$ is given by 
\begin{equation}
\delta v_{sing}={\sigma_c^2-\sigma^2\over 2\lambda_0}.
\label{deltapin}
\end{equation}
This form is shown as a dashed line in Fig.1 and fits very well our
data for $\sigma<\sigma_c$. This is easily understood because
in the region $\sigma<\sigma_c$ the flux line is localized onto the
strongest pin and its free energy is basically unaffected by the
presence of the other defects. 

The two sliding states of the driven CDW differ qualitatively in
the behavior of
the spatial fluctuations of the coarsed-grained phase $\phi({\bf r},t)$. 
For $\sigma<\sigma_c$ pinning dominates the
dynamics. The sliding state is rough with large spatial fluctuations
of the phase both in the directions parallel and perpendicular to the
external drive $\sigma$. For $\sigma>\sigma_c$ the term
$\sigma\partial_x\phi$ washes out the effect of pinning in the
$x$-direction, damping out phase fluctuations in this direction. In this
case, the spatial fluctuations of the phase are anisotropic and
are suppressed in the direction of the external drive.
This behavior is shown qualitatively in Fig. 3 that displays contour
plots of the CDW phase for increasing values of $\sigma$.

To quantify the behavior of phase fluctuations in the two sliding 
states, we have
examined the interface width $w(L_x,L_\perp)$ defined
in Eq. (\ref{width}). In the isotropic disorder-dominated phase for
$\sigma<\sigma_c$ we expect $w\sim L_x\sim L_\perp$.
To understand this, we recall that when
$\sigma=0$ the path-integral solution (\ref{pathint}) of the CDW
problem can also be interpreted as the transfer matrix solution of the
Schr\"odinger equation for a quantum particle in a random potential in
$d$ spatial dimensions and imaginary time\cite{hatano}. 
The width $w(L_x,L_\perp)$
corresponds to the fluctuations in the energy of the quantum particle
as a function of system size. For $d=1$ the quantum particle is always
localized. The states are exponentially localized 
and one can show \cite{nelsonnato} that the energy
fluctuations scale as the system size, i.e., $w(L_x)\sim L_x$. A similar
behavior is expected for $d=2$. In the
large-$\sigma$ phase, we postulate an anisotropic scaling ansatz for
the interface width,
\begin{equation}
w(L_x,L_\perp)=L_x^\chi f(L_\perp/L_x^z),
\label{aniso}
\end{equation}
where $\chi$ and $z$ are two new exponents. For $L_x^z>>L_\perp$ the
system looks one-dimensional, extended along the $x-$direction. An
approximate equation of motion in this regime is obtained from Eq.
(\ref{eomeffective}) with $\nabla\rightarrow\partial_x$. For large
$\sigma$ both pinning by impurities and intrinsic pinning which yields
the KPZ nonlinearity are negligible compared to the
$\sigma\partial_x\phi$ term and one can obtain an approximate solution
of the equation, which yields $w(L_x,L_\perp)\sim L_x^{1/2}$. 
\begin{figure}[hbt]
\epsfxsize=3.5in\epsfbox{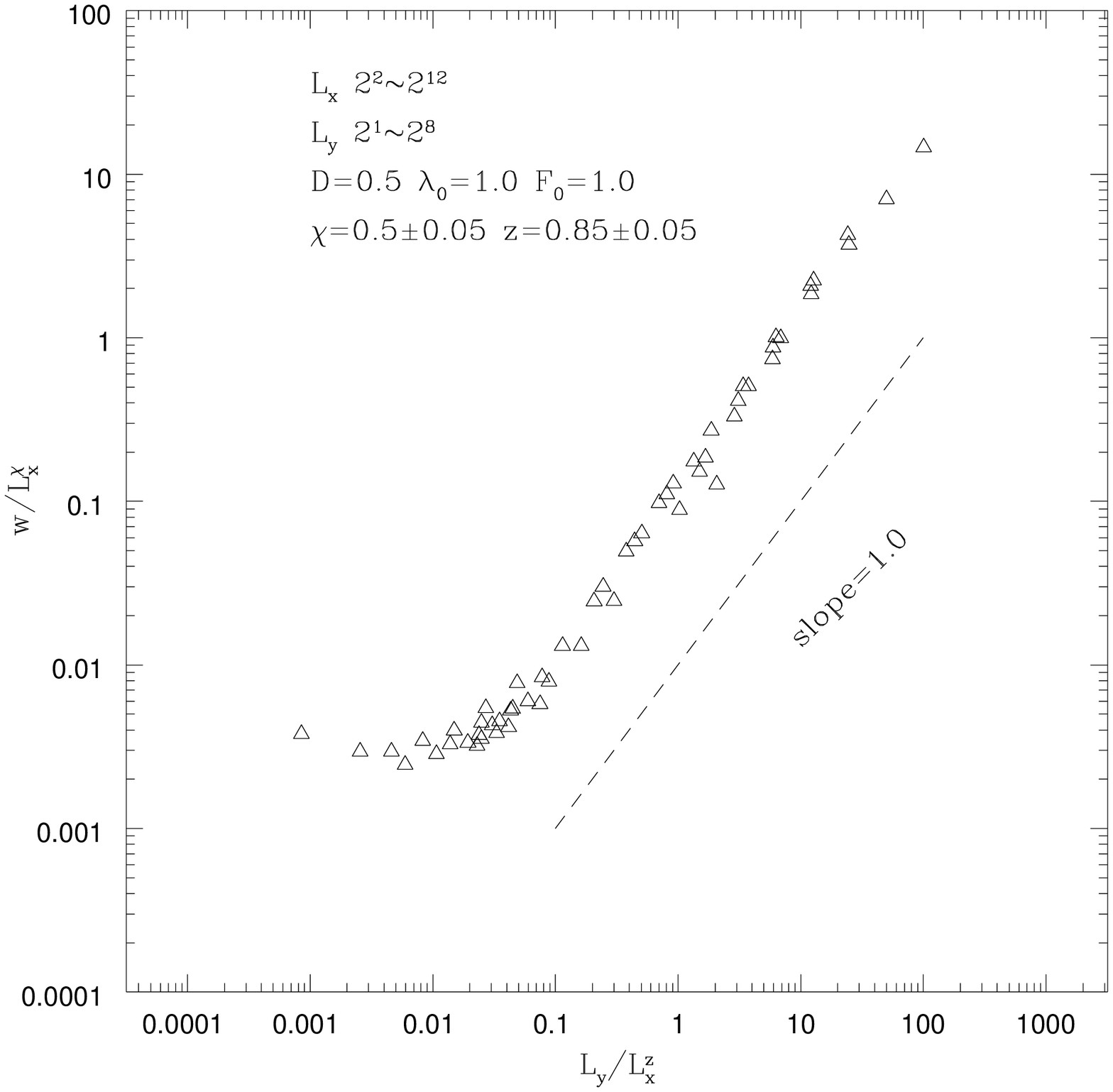}
\vspace{5pt}
\centerline{\bf Fig.4}
\end{figure}
The scaling function $f(s)$ in Eq. (\ref{aniso}) must therefore obey
$f(s)\sim s^{{\chi-1/2}\over z}$ for $s<<1$. This result is easily
understood by exploiting the mapping of driven CDW dynamics onto the
problem of the tilted magnetic flux line. For $\sigma>>\sigma_c$
$(h>>h_c)$ the flux line is delocalized and ``rides over'' many
columnar defects. The interface width $w(L_x,L_\perp)$ corresponds the
fluctuations of the flux line free energy, which in this limit is
determined by a sum of independent random energies, yielding a
$w(L_x,L_\perp)\sim L_x^{1/2}$ scaling for $L_x^z>>L_\perp$.
In the opposite limit, $L_\perp>>L_x^z$, the system looks
one-dimensional along the $y-$direction for $d=2$. The
$\sigma\partial_x\phi$ term has no effect and fluctuations are
dominated by disorder, with the result $w(L_x,L_\perp)\sim L_\perp$.
This follows again from the exponential localization of states of a
one-dimensional quantum particle in a random potential. 
The scaling function obeys $f(s)\sim s$ for $s>>1$ and
$w(L_x,L_\perp)\sim L_x^{\chi-z}L_\perp$ for $L_\perp>>L_x^z$ in
$d=2$. A scaling collapse of our numerical results for the interface
width in $d=2$ is shown in Fig. 4. The best collapse is achieved with
$\chi=0.5\pm 0.05$ and $z=0.85\pm 0.05$. These results agree well
with the asymptotic values discussed earlier.

\section{Conclusion}
We have studied the dynamics of driven CDWs moving through a random medium
at zero temperature, at driving forces well above the threshold force
where the depinning transition occurs. The CDW model considered incorporates
new nonequilibrium terms which are important in the strongly driven
regime and are generally not included in the 
FLR model. We have found that in $d=1,2$
the driven CDW exhibits a first order phase transition at a critical
driving force $\sigma_c\sim\sqrt{2\lambda_0F_0}$.
For $\sigma<\sigma_c$ disorder controls the dynamics
yielding a rough sliding phase with spatial fluctuations of the
phase that grow linearly with $L_x$, indicating that the phase-only
model breaks down.
For $\sigma>\sigma_c$, the driving force washes out the effect
of disorder in the direction of motion. The CDW slides uniformly with
$\delta v_{sing}=0$. This moving phase is highly anisotropic
as the external drive suppresses the spatial fluctuations of the phase
in this direction. The CDW remains ``rough'' in the direction
transverse to the external drive.
By using the Cole-Hopf transformation, the problem of 
CDW dynamics at large driving force can be mapped onto the problem of
the tilt response of a magnetic
flux-line pinned by columnar defects. 
The dynamical transition of the sliding CDW
corresponds to the transverse Meissner effect of the tilted flux line. 

Our coarsed-grained model of phase dynamics given in Eq.
(\ref{eomeffective}) only applies in the strongly driven phase, well
above the $T=0$ threshold field, $E_T$. 
In the weak pinning limit, one can relate $E_T$ to the Lee-Rice
length $L_0$ that represents the typical domain size in a pinned CDW.
The pinning length $L_0$ is the length where the elastic strains
induced by disorder are of order one and is given by
$L_0=\Big[{\hbar v_F\pi\over V_0}({\xi_0^{d/2}\over
c^{d-1}})\Big]^{2\over 4-d}$. Here $c$ is the average spacing between
the CDW chains and $\xi_0$ is the range of the disorder potential. We
expect $\xi_0\sim c$. The threshold
field is then estimated by balancing the total force on a domain of size
$L_0$ to the elastic force acting on the same domain, with the result,
$E_T={\hbar v_Fk_F\over 2ec^{d_\perp}}L_o^{d-2}$.
This corresponds to a threshold force $\sigma_T=E_Te({\tau\over m^*})$.
The first order transition
is predicted to occur at a critical force
\begin{equation}
\sigma_c\sim\sqrt{2\lambda_0F_0}, 
\label{criticalsigma}
\end{equation}
where to leading order in $1\over\sigma$, $\lambda_0\sim{W_0^2\over
8k_F^3\sigma}$ and $F_0\sim{V_0^2\over 4k_F\sigma}$. Recalling that
$\sigma=Ee\tau/m*$, we can solve Eq. (\ref{criticalsigma})
self-consistently for a critical field $E_c$, with the result
\begin{equation}
E_c=(e{\tau\over m^*})^{-1}\sigma_c=({\pi\rho_1\over e})\sqrt{W_0V_0}.
\label{eccdw}
\end{equation}
The first order transition will only be observable
if $E_c>>E_T$. We find
$E_c/E_T\sim{2\pi^2\rho_1\over k_F}\sqrt{W_0\over V_0}({\xi_0\over
L_o})$ for $d=2$.
The first order phase transition at $E_c$ may
therefore be observable in a dirty material ($L_0\approx c$) with
appreciable intrinsic pinning ($W_0>>V_0$). Using the results of our
calculation of the tilt response of a flux line pinned by a single defect
(Appendix A), we estimate the
magnitude of the jump discontinuity in $\delta v_{sing}$ at $\sigma_c$
as $|{d v_{sing}\over
d\sigma}|\approx{\sigma_c\over\lambda_0}=\sqrt{2F_0/\lambda_0}$. This 
corresponds to a jump
discontinuity in $G_{\rm diff}$ that can be expressed in terms of
microscopic CDW parameters
as $\Delta G_{\rm diff}=({k_F\over \pi c})({L_\perp\over L_x})
{2e^2\tau\over m*}({V_0\over
W_0})$ for $d=2$. The discontinuity in $G_{\rm
diff}$ is very small when the condition $W_0>>V_0$ of observibility
of transition is satisfied. 

\vspace{0.5in}
\centerline{\bf ACKNOWLEDGMENTS}
\vspace{0.5in}
We are grateful to the National Science Foundation for supporting
the work at Syracuse
University through Grant No. DMR92-17284 and DMR94-19257.
M.P.A.F. has been supported by grants PHY94-07194, DMR-9400142 and
DMR95-28578. L.-WC and MCM thank Alan
Middleton for many illuminating discussions.
\vspace{0.5in}

\appendix
\section{Single Flux Line and one columnar defect}

In order to gain some understanding of the dependence of our results on
the size $\Lx$ of the system in the direction of the external
driving force (or tilt field for the magnetic flux line),
it is instructive to consider the action of a transverse tilt
field on a flux line pinned
by a {\it single} attractive columnar defect.
The partition function is given by Eq. (\ref{partition}), with
$U({\bf r})$ the pinning potential due to the single impurity,
chosen for simplicity as an
attractive $\delta$-function,
\begin{equation}
\label{pot}
U(\bfr)=-U_0\delta(\bfr).
\end{equation}
Following Ref. \cite{nelsonnato},
Eq. (\ref{partition}) can be thought of as a path integral in
imaginary time and the partition function can be
rewritten as a quantum mechanical matrix
element,
\begin{equation}
\label{matrix}
Z(\bfr_\perp,{\bf 0};L)=<\bfr_\perp|e^{-L{\cal H}/T}|{\bf 0}>,
\end{equation}
where $|{\bf 0}>$ and $<\bfr_\perp|$ are initial and final states
localized at ${\bf 0}$ and $\bfr_\perp$, respectively,
and the ``Hamiltonian'' ${\cal H}$ is the operator,
\begin{equation}
\label{nonhham}
{\cal H}=-{1\over 2\tilde{\epsilon}_1}\big(T\grad-{\bf h}\big)^2+U(\bfr).
\end{equation}
The operator ${\cal H}$ is nonhermitian as
${\cal H}^\dagger(h) = {\cal H}(-h)$. To find its spectrum we need to solve
both the right and left eigenvalue problem, defined by,
\begin{equation}
\label{eigenR}
{\cal H}u^R_n=E_nu^R_n,
\end{equation}
and
\begin{equation}
\label{eigenL}
{\cal H}^\dagger u^L_n=E_nu^L_n,
\end{equation}
where $u^R_n(x)$ and $u^L_n(x)$ are the right and left eigenfunction,
respectively, normalized according to $\int_0^L dx u^L_n(x)u^R_n(x)=1$,
and $E_n$ are the corresponding eigenvalues.
The path integral  (\ref{partition}) can then be expressed in
terms of the eigenvalues and eigenfunctions of the fictitious
quantum problem,
as
\begin{equation}
\label{eigsum}
Z(\bfr_\perp,{\bf 0};L)=\sum_n u^R_n(\bfr_\perp)u^L_n({\bf 0})
e^{-E_nL/T}.
\end{equation}
This is also equivalent to writing  the path integral in terms of the
eigenvalues of a corresponding transfer matrix.
In the limit $L\rightarrow\infty$ the smallest eigenvalue dominates
and the free energy per unit length $g(h)$ of the flux line
is determined by the real part of the
ground state energy $E_0$ of the
quantum problem, according to $g(h)=E_0+h^2/2\tilde\epsilon_1$.
For localized states, the ground state wavefunction
$u_0(\bfr)$ determines the localization length of the flux line
\cite{nelsonnato}.

For simplicity we begin by considering a flux line in $1+1$ dimensions,
with $x$ the direction of the applied tilt field. The vortex free energy
is given by the ground state of the non-hermitian ``Schr\"odinger equation'',
given by\cite{groundenergy},
\begin{equation}
\Big[-{1\over 2\tilde{\epsilon}_1}\big(T{d\over dx}-h\big)^2-U_0\delta(x)
\Big]u(x)=Eu(x),
\label{shro}
\end{equation}
to be solved with periodic boundary conditions, $u(0)=u(\Lx)$.
We are considering here the right eigenvalue problem and dropping
for simplicity of notation the labels on the eigenfunction.
The solutions of Eq. (\ref{shro}) for $x\not=0$ are given by
$u_\pm(x)=e^{\pm\kappa x}e^{hx/T}$, with
$\kappa=\sqrt{-2\tilde{\epsilon}_1E/T^2}$. 
The general solution of Eq. (\ref{shro}) can then be written as
\begin{equation}
\label{eigf}
u(x)=Ae^{(h/T+\kappa)x}+Be^{(h/T-\kappa)x},
\end{equation}
where $A$ and $B$ are constants to be determined by the periodic boundary
condition and the condition that the wavefunction has a slope discontinuity
determined by the $\delta$-function,
$\big({du\over dx}\big)_{\Lx}-\big({du\over dx}\big)_0=
{2\tilde{\epsilon}_1U_0\over T^2}u(0)$.
The condition for a nontrivial solution to exists yields the
eigenvalue equation, given by,
\begin{equation}
\label{eigeq}
\cosh(h\Lx/T)-\cosh(\kappa\Lx)+{\tilde{\epsilon}_1U_0\over T^2\kappa}
\sinh(\kappa\Lx).
\end{equation}
In the limit $\Lx\rightarrow\infty$, there is one localized ground state
for $h<h_c$, with $h_c=\tilde{\epsilon}_1U_0/T$. For $h\geq h_c$ all states
are extended. The ground state energy is given by,
\begin{eqnarray}
\label{ground}
& & E_0^\infty=-{h_c^2\over 2\tilde{\epsilon}_1},\qquad h<h_c \nonumber \\
& & E_0^\infty=-{h^2\over 2\tilde{\epsilon}_1},\qquad h\geq h_c,
\end{eqnarray}
The ground state wavefunction is given by
\begin{eqnarray}
\label{groundwave}
& & u_0^R(x)=\sqrt{{T\over h_c}}e^{-(h_c-h)x\over T}
\qquad h<h_c \nonumber \\
& & u_0^R(x)={1\over\sqrt{\Lx}},\qquad h\geq h_c.
\end{eqnarray}%
It is exponentially localized for $h<h_c$, with
localization length $\xi\sim {T\over (h_c-h)}$.
If the system is infinitely long in the field$(z-)$ direction
$(L\rightarrow\infty)$, the flux line free energy per unit length is
simply $g(h)=E_0+h^2/2\tilde\epsilon_1$. The free energy per unit length
is shown in Fig.5 as a function of $h$ (thick line). Clearly the
magnetization $m_\perp$ is $m_\perp=-h/\tilde\epsilon_1$ below $h_c$
and cancels the applied field, yielding $b_\perp=0$, as required for 
transverse Meissner effect. For $h\geq h_c$,
$b_\perp=h/\tilde{\epsilon}_1$, which is the value in the absence
of disorder. The induction has a jump discontinuity at $h_c$.

We now discuss the corrections to the above results due to a finite
system size $\Lx$. This will be useful for understanding our
numerical results for the driven CDW. One can study analytically the
finite size corrections in the limit $L_xh>>1$. Keeping the leading
finite size correction, the real part of the ground state energy 
is given by
\begin{eqnarray}
E_0{(L_x)} & \approx & -{h_c^2\over
2\tilde\epsilon_1}\Big[1+2e^{-L_x(h_c-h)/T}\Big],\qquad h<h_c\nonumber
\\
E_0{(L_x)} & \approx &  -{h^2\over 2\tilde\epsilon_1}+{1\over
L_x}\Big({hT\over\tilde\epsilon_1}\Big)ln\Big(1-{h_c\over h}\Big)\nonumber \\
& \approx & -{h^2\over 2\tilde\epsilon_1}-{Th_c\over
L_x\tilde\epsilon_1}
,\qquad\qquad\qquad h\geq h_c,
\label{finitele}
\end{eqnarray}
where we have assumed $h>>h_c$. In the region $h<h_c$, where the flux
line is pinned on the defect, the finite size corrections vanish
exponentially as $L_x\over\xi$, with $\xi$ the localization length.
For $h>h_c$ the flux line is depinned and samples the whole sample. In
this region the asymptotic $(h>>h_c)$ correction to the ground state
energy of an infinite system vanishes only as $1\over L_x$. The
free energy $g(h,L_x)$ is shown in Fig. 5 for a few values of system
size $L_x$. 

The CDW frequency shift $\delta v_{sing}$ is simply
propotional to the negative of $g(h)$ (see table 1). The similarity in
the finite size dependence of Figs. 1 and 5 is apparent.
Finally, Fig. 6 shows the dependence of the normalized probability
density in the ground state $P(x)=u^R(x)u^L(x)$
on system size $L_x$. The tilt induces a mismatch of order
$e^{-(h_c-h)L_x/T}$ in the wavefunction at the system boundaries. For
$h<<h_c$ this mismatch is exponentially small and the periodic
boundary conditions are easily accommodated. For $h\sim h_c$ the
boundary conditions force a large negative gradient in the wave
function, as apparent from Fig. 6. Physically this corresponds to the fact
that for $h\sim h_c$ and finite $L_x$ long sections of the flux line
are still pinned on a strong columnar pin, yielding large finite size
effect in the region above the depinning transition.

\begin{figure}[hbt]
\epsfxsize=3.5in\epsfbox{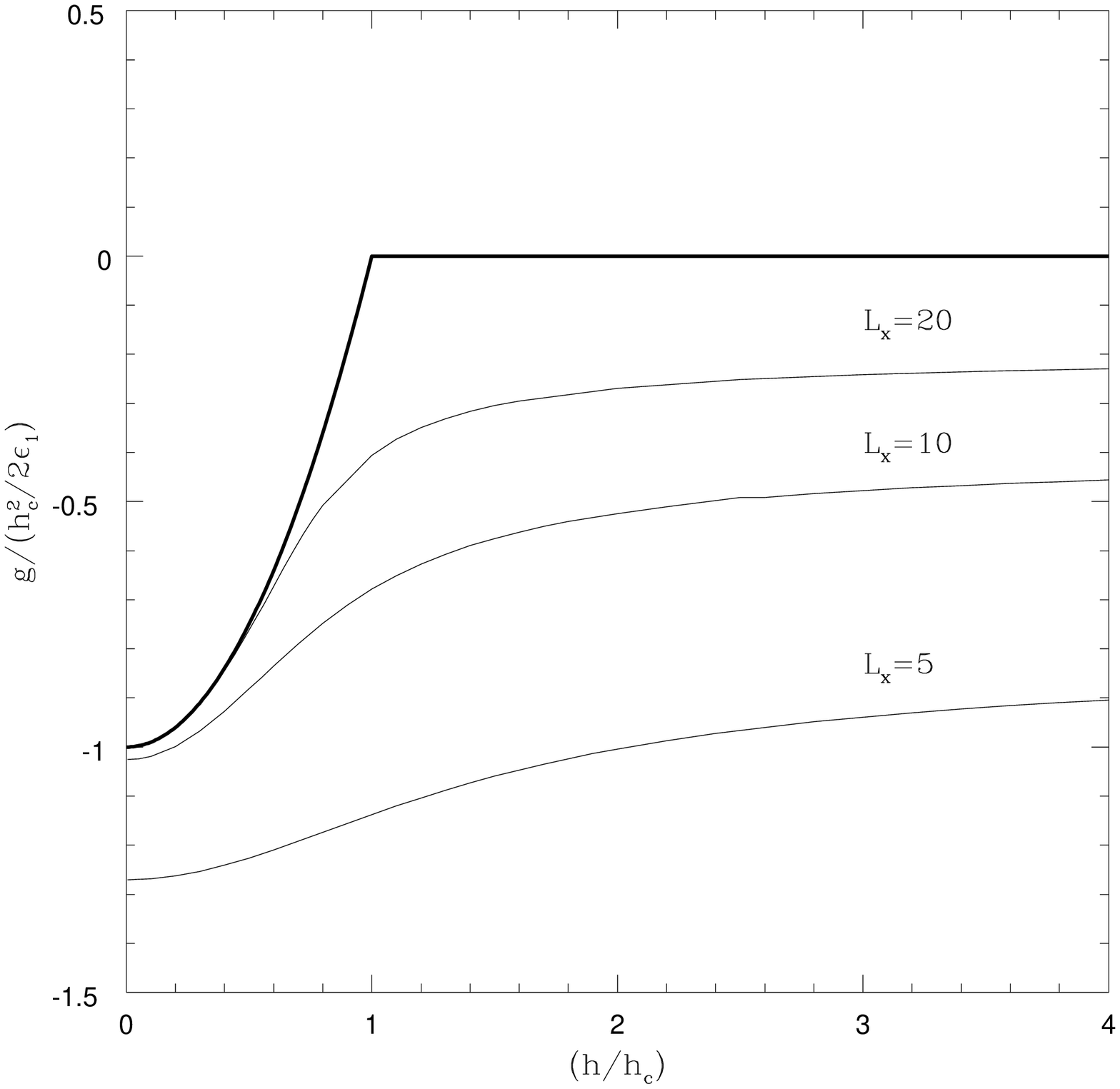}
\vspace{5pt}
\centerline{\bf Fig.5}
\end{figure}

\begin{figure}[hbt]
\epsfxsize=3.5in\epsfbox{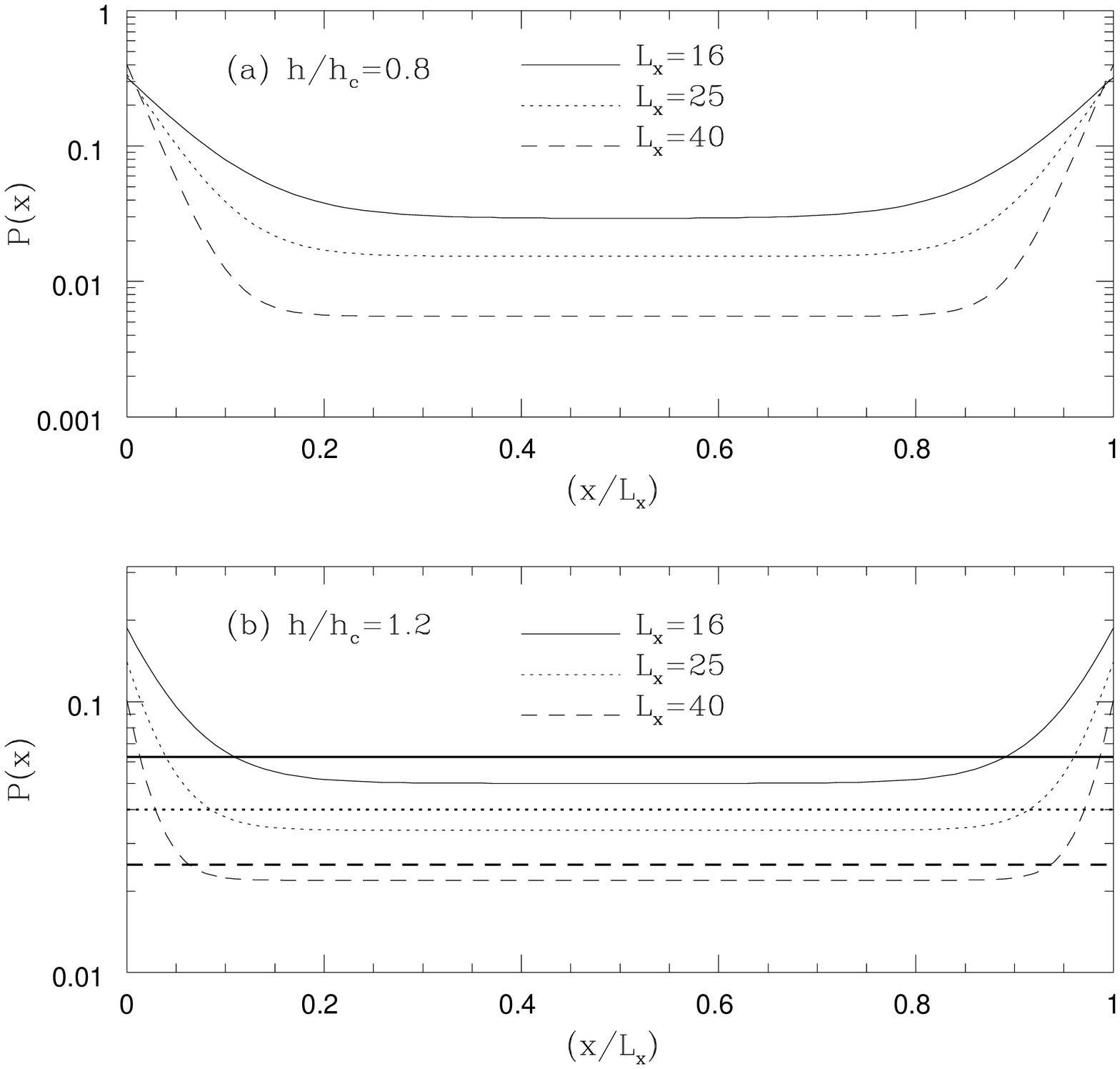}
\vspace{5pt}
\centerline{\bf Fig.6}
\end{figure}

\pagebreak[4]
\centerline{\bf Figure Captions}
\vspace{0.5in}
\noindent
{\bf Fig. 1}\hspace{0.25in}
The singular part $\delta v_{sing}$ of the CDW velocity 
as a function of applied force $\sigma$
for various system sizes. The critical force $\sigma_c$ is estimated to be
$\sigma_c\sim\sqrt{2\lambda_0F_0}=\sqrt 2$ for the set of parameters 
used in the figure.

\noindent {\bf Fig. 2}\hspace{0.25in}
The derivative of the CDW velocity $<\tphi>$ with respect to the
applied forces $\sigma$. This figure can also be interpreted as the
transverse magnetization $m$(measured in units of $\phi_0\over 4\pi$)
versus transverse magnetic field $H_x$
for a flux line in columnar defects.

\noindent {\bf Fig. 3}\hspace{0.25in}
Two-dimensional CDW phase configurations $\phi(x,y;t)-\overline{\phi(x,y;t)}$  
for various driving forces $\sigma$ at long time for $\lambda_0=1.0$
and $F_0=1.0$. The estimated $\sigma_c$ is $\sigma_c=\sqrt 2$. The
contour plots are (a)$\sigma=0.1$, (b)$\sigma=1.4$ and (c)$\sigma=10.0$. 
The relative value of the CDW
phase $\phi$ is given by the greyscale intensity, with the
brightest spots corresponding to the highest $\phi$ and the darkest spots
the lowest $\phi$.  
 
\noindent {\bf Fig. 4}\hspace{0.25in}
The figure shows the scaling collapse of the interface width
$w(L_x,L_\perp)$ according to the ansatz Eq. (\ref{aniso}) for $d=2$.
The parameter values are indicated and $\sigma=50.0$.

\noindent {\bf Fig. 5}\hspace{0.25in}
The free energy per unit length $g(h)$ of a tilted flux line pinned by
a single defect as a function of $h\over\tilde\epsilon_1$. Both the
result in the thermodynamic limit $L_x\rightarrow\infty$ (thick line) 
and the finite size results are shown.

\noindent {\bf Fig. 6}\hspace{0.25in}
Probability density $P(x)=u^R(x)u^L(x)$ for various system sizes at
(a) $h=0.8h_c$ and (b) $h=1.2h_c$. In Fig. (a) the probability density
decays exponentially to zero over a length $x\sim\xi\sim 5$ in the
limit $L_x\rightarrow\infty$. The finite-size correction are
determined by the value of $P(x)$ in the central flat region. They are
therefore rather small even for not too large a value of $L_x$. In
Fig. (b) $P(x)={1\over L_x}$ in the limit $L_x\rightarrow\infty$. This
value is shown as an horizontal line for each $L_x$.

\end{multicols}

\begin{references}
\bibitem{gruner}
G. Gruner, Rev. Mod. Phys. {\bf 60}, 1129(1988).
%
\bibitem{fisherhuse}
D. S. Fisher, M. P. A. Fisher, D. Huse, Phys. Rev. {\bf B43}, 130(1991).
%
\bibitem{cdwsimul}
P. Sibani and P. B. Littlewood, Phys. Rev. Lett. {\bf 64}, 1305(1990).
1336(1991).
%
\bibitem{cdwnarayan}
O. Narayan and D. S. Fisher, Phys. Rev. {\bf B 46}, 11520(1992).
{\bf 263}, 943(1994).
%
\bibitem{vinokurmoving}
A. E. Koshelev and V. M. Vinokur, Phys. Rev. Lett. {\bf 73}, 3580(1994).
%
\bibitem{balents}
L. Balents and M. P. A. Fisher, Phys. Rev. Lett. {\bf 75}, 4270(1995).
%
\bibitem{DSFcdw}
L. Sneddon, M. C. Cross and D. S. Fisher, Phys. Rev. Lett. {\bf 49},
292(1982).
%
\bibitem{unpublished} L.-W. Chen, L. Balents, M. P. A. Fisher, and
M. C. Marchetti, unpublished.
%
\bibitem{kpz}
M. Kardar, G. Parisi, and Y. C. Zhang, Phys. Rev. Lett {\bf 56},
889(1986).
%
\bibitem{flr}
H. Fukuyama and P. A. Rice, Phys. Rev. {\bf B17}, 535(1978).
%
\bibitem{krug}
J. Krug, Phys. Rev. Lett {\bf 75}, 1795(1995).
%
\bibitem{sinegordonrg}
M. Rost and H. Spohn, Phys. rev. {\bf E49}, 3709(1994).
%
\bibitem{family}
J. G. Amar, P.-M. Lam, and F. Family, Phys. Rev. {\bf E47},
3242(1993).
%
\bibitem{etras}
D. Ertas and M. Kardar, Phys. Rev. {\bf E48}, 1228(1993).
%
\bibitem{balentssimon}
L. Balents, S. H. Simon, Phys. Rev. {\bf B51}, 6515(1995).
%
\bibitem{hatano}
N. Hatano and D. R. Nelson, cond-mat/9603165.
%
\bibitem{nelsonnato}
See, for example, D. R. Nelson in {\it Phenomenology and Applications
of High Temperature Superconductors: the Los Alamos Symposium, 1991},
(Addison-Wesley, Mass. 1992).
%
\bibitem{groundenergy}
As discussed in Ref.\cite{hatano}, the eigenvalues of the
non-hermitian operator $\cal H$ are in general complex. Here we
discuss in detail only the real part of the ground state eigenvalue
which determines the flux line free energy. The ground state
eigenvalue is real for $L_x\rightarrow\infty$, but acquires a nonzero
imaginary part at finite $L_x$ for $h>h_c$.
\end{references}
\end{document}